\documentclass{article}

\usepackage{latexsym}
\usepackage{epsf,axodraw,graphicx}

\begin{document}

\title{\bf Factorization of the Two Loop Four-Particle
Amplitude in Superstring Theory Revisited}

\author{
Zhi-Guang Xiao\footnote{E-mail: xiaozhiguang@pku.edu.cn}
 \\ School of Physics, Peking University \\
Beijing 100871, P. R. China\\ \\
Chuan-Jie Zhu\footnote{E-mail: zhucj@itp.ac.cn} \thanks{Supported
in part by fund from the National Natural Science Foundation of
China with grant Number
90103004 and 10475104.} \\
Institute of Theoretical Physics,
Chinese Academy of Sciences\\
P. O. Box 2735, Beijing 100080, P. R. China}

\maketitle

\abstract{ We study in detail the factorization of the newly
obtained two-loop four-particle amplitude in superstring theory.
In particular some missing factors from the scalar correlators are
obtained correctly, in comparing with a previous study of the
factorization in two-loop superstring theory. Some details for the
calculation of the factorization of the kinematic factor are also
presented. }

\newpage

Explicit result  for higher loop amplitudes in superstring is
quite rare. To our knowledge the only explicitly known higher loop
($\ge 2$) non-vanishing amplitude is the four-particle amplitude
in superstring theory, firstly obtained in \cite{IengoZhu} and
later re-obtained in \cite{WuZhu1,WuZhu2} in an explicitly gauge
independent way, following the works of D'~Hoker and Phong
\cite{DHokerPhonga,DHokerPhongb,DHokerPhongc,DHokerPhongd,
DHokerPhonge} on two loop measure of superstring theory. Recently
D'~Hoker and Phong \cite{DHokerPhongaa,DHokerPhongbb} also gave a
measure for three loop superstring theory. It  remains to see if
this can be used to do explicit three loop computations in
superstring theory. For other promising approach of covariant
calculation of superstring amplitudes we refer the reader to
Berkovits's review \cite{Berkovits}.

Due to the rareness of explicit results, it is natural to study
the known result in depth. The old result was casted into an
explicit modular invariant form \cite{IengoZhu2} and used in
\cite{Iengo} to prove the vanishing of the $R^4$ correction
\cite{GrossWitten,Green}. It has also been proved in \cite{WuZhu3}
that the results obtained in \cite{IengoZhu,WuZhu1,WuZhu2} are
equivalent. Another goal we have in mind is to make connection
with known results from field theory in ${\cal N}=4$
supersymmetric Yang-Mills theory \cite{Berna,Bernb}. This seems a
trivial problem, but in fact it turns out to be quite tricky. One
could use unitarity to fix the overall factor for the
four-particle amplitude.  In order to do this we need to know the
precise overall factor for other one loop amplitudes involved. Due
to the incomplete results in the literature, we therefore computed
all the relevant amplitudes in a consistent way and fixed all the
overall factors by either using factorization or unitarity. We
will present a detailed study of factorization and unitarity in
superstring theory in a later publication \cite{XiaoZhu}. In this
paper we present only the result of factorization of the two loop
four-particle amplitude in superstring theory.

The factorization of the two loop four-particle amplitude in
superstring theory was studied before in \cite{Zhu,Yasuda}.
Compared with these studies, we improved on obtaining some missing
factors from the scalar correlators. Also some details for the
calculation of the factorization of the kinematic factor are
presented. We will need these results in the forthcoming paper
\cite{XiaoZhu}.

To begin with, let us recall the two-loop four-particle amplitude
in type II superstring theories obtained in
refs.~\cite{WuZhu1,WuZhu2}:
\begin{eqnarray}
\mathcal {A}_{II} & = & C_{II}(2\pi)^{10}\delta^{10}
 (\sum_i k_i)g_c^4\, K(k_i,\epsilon_i)\,
 \int {1\over T^5}
{\prod _{i=1}^6{\rm d}^2a_i\over {\rm d}
V_{pr}|\prod_{i<j}a_{ij}|^2} \nonumber \\
& & \times \prod_{i=1}^4 {{\rm d}^2z_i\over|y(z_i)|^2} \prod_{i<j}
 \exp\{-k_i\cdot k_j \langle X(z_i)X(z_j)\rangle
\},\nonumber \\
& & \times
\Big|s(z_1z_2+z_3z_4)+t(z_1z_4+z_2z_3)+u(z_1z_3+z_2z_4)\Big|^2 ,
\end{eqnarray}
where
\begin{eqnarray}
{\rm d} V_{pr} & = & {{\rm d}^2 a_i {\rm d}^2 a_j {\rm d}^2 a_k
\over
|a_{ij}a_{ik}a_{jk}|^2}, \\
 T & = & \int {\rm d}^2 z_1{\rm d}^2z_2
\frac{|z_1-z_2|^2}{|y(z_1)y(z_2)|^2}, \\
y^2(z) & = & \prod_{i=1}^6\, (z-a_i)\, ,
\end{eqnarray}
and $\langle X(z_i)X(z_j)\rangle\equiv \langle X(z_i,\bar{z}_i)
X(z_j,\bar{z}_j)\rangle$'s are the scalar correlators (see below
in eq.~(\ref{correlator}) for explicit formulas in terms of prime
form and holomorphic differentials). The $K(k_i,\epsilon_i)$ is
the standard kinematic factor appearing at tree, one- and two-loop
computations \cite{GreenSchwarz,IengoZhu,WuZhu2}. $C_{II}$ is an
overall factor which should be determined from factorization and
unitarity \cite{XiaoZhu}.

In the dividing degeneration limit: $a_2-a_1=u$, $a_3-a_1=vu$,
$u\to 0$, we have
\begin{eqnarray} T\to
 \frac{2}{|u|}T_1T_2
 \frac1{|a_{14}a_{15}a_{16}|}
\end{eqnarray}
where
\begin{eqnarray}
 T_i & = & \int {{\rm d}^2z\over |y_i(z)|^2},
\\
 y_1(z) & = & \sqrt{z(z-1)(z-v)},
\\
 y_2(z) & = & \sqrt{(z-a_1)(z-a_4)(z-a_5)(z-a_6)} .
\end{eqnarray}

Setting
 \begin{equation}
 z_1=x_1 u+a_1, \qquad \hbox{and} \qquad
 z_2=x_2 u+a_1,
\end{equation}
we have:
\begin{eqnarray}
{ |\prod_{i<j}a_{ij}|^2
 \over
 |a_{45}a_{46}a_{56}|^2 }
 & \to & |u|^6|v(1-v)|^2
 |a_{14}a_{15}a_{16}|^6 , \\
|y(z_1)|^2
 & \to &
 |u|^3|y_1(x_1)|^2|a_{14}a_{15}a_{16}| ,
\\
|y(z_4)|^2
 & \to &
 |(z_4-a_1)|^2|y_2(z_4)|^2, \\
\prod_{i=1}^4{|y(z_i)|^2}
 & \to &
 |u|^6|a_{14}a_{15}a_{16}|^2 |(z_3-a_1)(z_4-a_1)|^2
 \nonumber \\
 & & \times
 |y_1(x_1)y_1(x_2)y_2(z_3)y_2(z_4)|^2 ,
\\
{\rm d}^2a_1{\rm d}^2 a_2{\rm d}^2 a_3
 & \to &
 |u|^2{\rm d}^2 u{\rm d}^2 v {\rm d}^2 a_1, \\
{\rm d}^2 z_1 {\rm d}^2 z_2 & \to &
 |u|^4 {\rm d}^2 x_1{\rm d}^2 x_2, \\
 & & \hskip -3cm
 \Big|s(z_1z_2+z_3z_4)+t(z_1z_4+z_2z_3)+u(z_1z_3+z_2z_4)\Big|^2
 \nonumber \\
 & \to
 & |s|^2|(z_3-a_1)(z_4-a_1)|^2 .
\end{eqnarray}

By using the above results we have
\begin{eqnarray}
 \mathcal{A}_{II}
& \to &
 C_{II}(2\pi)^{10}\delta^{10}(\sum_i k_i)g_c^4K(k_i,\epsilon_i)
 \nonumber \\
 & & \times \int
 {{\rm d}^2 u\over |u|2^5T_1^5T_2^5}
 {{\rm d}^2 a_1\over|a_{14}a_{15}a_{16}|^3}
 {{\rm d}^2v\over|v(v-1)|^2}\
\nonumber \\
& &   \times
 {{\rm d}^2x_1{\rm d}^2x_2\over|y_1(x_1)y_1(x_2)|^2}
 {{\rm d}^2z_3{\rm d}^2z_4\over|y_2(z_3)y_2(z_4)|^2}
\nonumber \\
& &   \times s^2\prod_{i<j}
 \exp\{-k_i\cdot k_j \langle X(z_i) X(z_j) \rangle\} .
 \label{factorizea}
\end{eqnarray}
In order to prove that the above amplitude factorizes correctly,
we need to study the degeneration limit of the scalar correlators.

The expression in the exponential factor in eq.~(\ref{factorizea})
is:
\begin{eqnarray}
 -\sum_{i<j}
 k_i\cdot k_j\langle X(z_i)X(z_j)\rangle
& = & \frac{s}2 \Big( \langle X(z_1)X(z_2)\rangle+\langle
X(z_3)X(z_4)\rangle \nonumber
\\
& & \hskip -.5cm - \langle X(z_1)X(z_3)\rangle-\langle
X(z_2)X(z_4)\rangle \Big)
\nonumber \\
& & \hskip -.5cm + \frac{t}2 \Big( \langle
X(z_1)X(z_4)\rangle+\langle
X(z_2)X(z_3)\rangle \nonumber \\
& & \hskip -.5cm - \langle X(z_1)X(z_3)\rangle-\langle
X(z_2)X(z_4)\rangle \Big) .
\end{eqnarray}
where the scalar correlator $\langle X(z_i)X(z_j) \rangle$ is
given as follows:
\begin{eqnarray}
 \langle X(z_i)X(z_j) \rangle   & = & -\ln(|E(z_i,z_j)|^2)
 \nonumber \\
 & & +2\pi({\rm Im} \int_{z_i}^{z_j}
 \omega)({\rm Im} \tau)^{-1}({\rm Im} \int_{z_i}^{z_j}\omega) ,
 \label{correlator}
 \end{eqnarray}
in terms of the prime form $E(z_i,z_j)$, the period matrix:
\begin{equation}
 \tau = \left(\matrix{\tau_{11}
 & \tau_{12}\cr \tau_{21} & \tau_{22}}\right),
 \end{equation}
and the holomorphic differentials:
\begin{equation}
 \omega(z) = \left(\matrix{\omega_1(z) \cr
 \omega_2(z)}\right) .
\end{equation}


\begin{figure}
\centerline{\includegraphics[width=9cm] {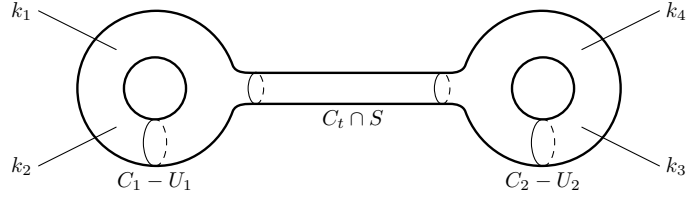}}\caption{The
dividing
 degeneration limit of a genus 2 surface.}
 \end{figure}

In the degeneration limit $u=t^4\to 0$ (see Fig.~1), we have the
following explicit formulas from Fay's book \cite{Fay} (page 38,
Proposition 3.1 and page 41, Corollary 3.2):
\begin{eqnarray}
\omega_1(x,t)
 & = & \left\{ \begin{array}[]{ll} v_1(x)+
 \frac{t}{4}v_1(p_1)u_1(x,p_1)+O(t^2), &(x\in C_1-U_1)
 \cr
 \frac{t}{4}v_1(p_1)u_2(x,p_2)+O(t^2), &(x\in C_2-U_2)
 \end{array}\right. , \nonumber \\
 \omega_2(x,t)
& = & \left\{
 \begin{array}[]{ll}
 \frac{t}{4}v_2(p_2)u_1(x,p_1)+O(t^2), &(x\in C_1-U_1)
\cr
 v_2(x)+\frac{t}{4}v_2(p_2)u_2(x,p_2)+O(t^2),
 &(x\in C_2-U_2)
\end{array}
 \right. , \nonumber
 \end{eqnarray}
and
\begin{equation}
 \tau \to
 \left(\matrix{\tau_1& 0\cr 0&\tau_2}\right )+
 2\pi i\frac{t}{4}\, \left(\matrix{ v_1^2(p_1)& v_1(p_1)v_2(p_2)\cr
 v_1(p_1)v_2(p_2)& v_2^2(p_2)}\right)+O(t^2) .
\end{equation}
Here $v_i(z)$ is the holomorphic differential on the $i$-th torus
(see below in eq.~(\ref{diffa}) for explicit expression). $p_1$ is
the $\infty$-point on the first torus ($p_1=\infty$) and $p_2$ is
one of the branch point on the second torus ($p_2=a_1$). The
Abelian differential of the second kind $u_i$ on the two tori are
given as follows:
\begin{eqnarray}
u_i(x,y) & = & -v_i(x)v_i(y){\partial^2\ln\vartheta\over \partial^2
z}(e_i) , \\
e_i & = & \int_x^y v_i-\frac1{2}(\tau_i+1) .
\end{eqnarray}
By using the above results and the degeneration property of the
theta function, the correlators are degenerated as follows:
\begin{eqnarray}
 \langle X(z_1)X(z_2)\rangle &\to &-G_1(x_1,x_2)
 +\ln|v_1(x_1)v_1(x_2)|-\ln|u|^2 , \label{eq25}
\\
 \langle X(z_3)X(z_4)\rangle &\to &-G_2(z_3,z_4)
 +\ln|v_2(z_3)v_2(z_4)| ,
\\
 \langle X(z_1)X(z_3)\rangle &\to &-G_1(x_1,p_1)-G_2(p_2,z_3)
 +\ln|v_1(x_1)v_2(z_3)| \nonumber \\
& & -\ln|K_1K_2/4|
 -\ln|u|^{1/2}-\ln|a_{14}a_{15}a_{16}|^{1/2} ,
 \\
 \langle X(z_2)X(z_4)\rangle &\to &
 -G_1(x_2,p_1)-G_2(p_2,z_4)+\ln|v_1(x_2)v_2(z_4)| \nonumber \\
& & -\ln|K_1K_2/4|
 -\ln|u|^{1/2}-\ln|a_{14}a_{15}a_{16}|^{1/2} , \label{eq28}
 \end{eqnarray}
where we have defined:
\begin{eqnarray}
G_i(x,y)=\ln\Big|{\vartheta_1(\int_{x}^{y}v_i,\tau_i)
 \over \partial\vartheta_1(0,\tau_i)}\Big|^2
 -2\pi ({\rm Im}\int_{x}^{y}v_i)({\rm Im}
 \tau_i)^{-1}({\rm Im}\int_{x}^{y}v_i),
\end{eqnarray}
and
\begin{eqnarray}
 v_1(z) & = & {{\rm d} x\over K_1 y_1(x)} = {\rm d} \tilde x,
\hspace{0.5cm}
 v_2(z)={{\rm d} z\over K_2 y_2(z)} = {\rm d} \tilde z,  \label{diffa} \\
 K_1 & = & \oint_{\alpha_1} {{\rm d} x\over   y_1(x)}   ,
\hspace{0.5cm}
 K_2=\oint_{\alpha_2}  {{\rm d} z\over   y_2(z)}   .
\end{eqnarray}
By using these results we have:
\begin{eqnarray} & & \hskip -2cm -\sum_{i<j}
 k_i\cdot k_j\langle X(z_i)X(z_j)\rangle \nonumber \\
&\to &
 \frac{s}{2}
 \Big[ -G_1(x_1,x_2)-G_2(z_3,z_4)+G_1(x_1,p_1) \nonumber \\
 & &
 +G_2(p_2,z_3)+G_1(x_2,p_1)+G_2(p_2,z_4)
\nonumber \\
& & +\ln\big( |a_{14}a_{15}a_{16}||K_1K_2/4|^2\big) \Big]
+\ln|u|^{-s/2},
\end{eqnarray}
and so the two loop four-boson amplitude becomes:
\begin{eqnarray}\label{4PointsFact}
 \mathcal{A}_{II}
& \to &
 iC_{II}(2\pi)^{10}\delta^{10}(\sum_i k_i)g_c^4 \,
 K(k_i, \epsilon_i)
\nonumber \\
& &  \int
 {{\rm d}^2 u\over 2^5T_1^5T_2^5}
 {{\rm d}^2 a_1\over|a_{14}a_{15}a_{16}|^3}
 {{\rm d}^2v\over|v(v-1)|^2} \nonumber \\
& & \times
 {{\rm d}^2x_1{\rm d}^2x_2\over|y_1(x_1)y_1(x_2)|^2}
 {{\rm d}^2z_3{\rm d}^2z_4\over|y_2(z_3)y_2(z_4)|^2}
\nonumber \\
& & \times
 \exp\Big\{
\frac{s}{2} \Big( -G_1(x_1,x_2)-G_2(z_3,z_4)+G_1(x_1,p_1)
\nonumber \\
& & \qquad +G_2(p_2,z_3)+G_1(x_2,p_1)+G_2(p_2,z_4)
 \Big)
 \Big\}
\nonumber \\
& & \times
 s^2|a_{14}a_{15}a_{16}|^{s/2}|K_1K_2/4|^s|u|^{-s/2-1}
\nonumber \\
&\to &
 iC_{II}(2\pi)^{10}\delta^{10}
 (\sum_i k_i)g_c^4 \, K(k_i, \epsilon_i)
\nonumber \\
& &
 \times {16\pi\over-s+2}
 \int  {1\over 2^5T_1^5T_2^5}
 {{\rm d}^2 a_1\over|a_{14}a_{15}a_{16}|^2}
 {{\rm d}^2v\over|v(v-1)|^2} \nonumber \\
& & \times  {{\rm d}^2x_1{\rm d}^2x_2\over|y_1(x_1)y_1(x_2)|^2}
 {{\rm d}^2z_3{\rm d}^2z_4\over|y_2(z_3)y_2(z_4)|^2} \times
  | K_1K_2/4|^2 \nonumber \\
& &\times  \exp\Big\{   -
 \Big(  G_1(x_1,x_2)- G_1(x_1,p_1) - G_1(x_2,p_1)\Big)
 \nonumber \\
& & \qquad -\Big( G_2(z_3,z_4)  - G_2(z_3,p_2) - G_2(z_4, p_2)
 \Big)  \Big\}\,  . \label{factorc}
\end{eqnarray}

In order to prove that the above amplitude do factorize correctly,
we need to do two things. The first is to prove that the kinematic
factor $K(k_i,\epsilon_i)$ factorizes into a summation of two
one-loop 3-particle kinematic factor. The summation is over all
possible intermediate states. The second is to prove that the
rests in eq.~(\ref{factorc}) are the product of two one-loop
3-particle amplitudes apart from the propagator $\frac{1}{-s+2}$.
Let us study the second first.

The first non-vanishing contributions is from a massive
intermediate  states. Only the  massive boson from the NS-NS
sector contributes because of the super-ghost. So we need to
compute the 1-loop amplitude for  two massless bosons to one
massive boson. The massive boson vertex operator in the
$0$-picture \cite{IengoMarisa,MassiveVertexa,MassiveVertexb} is:
\begin{eqnarray}
& & \hskip -1cm {\cal V}^{(0,0)}_{_M}
(k,\alpha_{\mu\nu\rho},\sigma_{\mu\nu},
 \sigma_\mu) \nonumber \\
 & = & -g_{_M}\Big\{\alpha_{[\mu\nu\rho]}\big[ 3 i\partial
 X^\mu\psi^\nu\psi^\rho - k\cdot\psi
 \psi^\mu\psi^\nu\psi^\rho\big] \nonumber \\
 & & +\sigma_{\mu\nu} \big[i \partial X^\mu\partial
 X^\nu+\partial \psi^\mu \psi^\nu+k\cdot\psi i\partial X^\mu
 \psi^\nu\big] \nonumber \\
 & & +\sigma_\mu\big[i
 \partial^2 X+k\cdot\psi \partial
 \psi^\mu\big] \Big\}\Big\{ \mbox{right-moving part}\Big\}
\, e^{ik\cdot X} ,
\end{eqnarray}
where $\alpha_{\mu\nu\rho}$, $\sigma_{\mu\nu}$ and $\sigma_\mu$
are the polarization tensors and we set the following
normalization:
\begin{equation}
\alpha_{\mu\nu\rho}(k)\alpha^{\mu\nu\rho}(-k) = -{1\over 6},
\qquad \sigma_{\mu\nu}(k)\sigma^{\mu\nu}(-k) = 1 .
\end{equation}
By convention $\alpha(-k) = \alpha^*(k)$ and
$\sigma(-k)=\sigma^*(k)$. For later use we give here also the
completeness relations for these tensors:
\begin{eqnarray}
\sum_\alpha
\alpha_{\mu_1\nu_1\rho_1}(k)\alpha_{\mu_2\nu_2\rho_2}(-k)
 & = & -\frac{1}{288} \Big( \big[ ({k}_{\mu_1}{k}_{\mu_2} +
 2 g^{\mu_1 \mu_2}) \nonumber \\
 & &   \times
({k}_{\nu_1}{k}_{\nu_2} + 2 g^{\nu_1 \nu_2})
({k}_{\rho_1}{k}_{\rho_2} + 2 g^{\rho_1 \rho_2}) \nonumber \\
& & \hskip -1.5cm - (\mu_1\leftrightarrow \nu_1) -
(\mu_1\leftrightarrow \rho_1) \big] - ( \nu_1\leftrightarrow
\rho_1) \Big), \label{completealpha}
\end{eqnarray}
\begin{eqnarray}
\sum_\sigma \sigma^{\mu_1\mu_2}(k)\sigma^{\nu_1\nu_2}(-k) & = &
\frac{1}{8}[(k^{\mu_1}k^{\nu_1} + 2 {\eta^{\mu_1\nu_1}})
(k^{\mu_2}k^{\nu_2} + 2 {\eta^{\mu_2\nu_2}}) \nonumber \\
& & \hskip -2.5cm - {1\over 9} (k^{\mu_1}k^{\mu_2} + 2
{\eta^{\mu_1\mu_2}}) (k^{\nu_1}k^{\nu_2} + 2 {\eta^{\nu_1\nu_2}})
+ (\mu_1 \leftrightarrow \mu_2) ] . \label{completesigma}
\end{eqnarray}

The (bosonic) $n$-particle one loop amplitude are obtained as
follows:
\begin{eqnarray} \mathcal{A}^{1-loop}_n=\int {\rm d}\mu
\prod_{i=1}^n{\rm d}^2z_i \sum_s \eta_s
Q_s\big\langle\prod_{i=1}^n
\mathcal{V}(k_i,\epsilon_i,z_i)\big\rangle_s,
\end{eqnarray}
and for $n=3$ (two massless ($k_1$ and $k_2$) and one massive ($
k_3 =k$)), we have
\begin{eqnarray}
 & & \hskip -1cm \mathcal{A}^{\rm
1-loop}_{3} \nonumber \\
&=& -ig_c^2g_{_M}C_I(2\pi)^{10} \delta^{10}(k_1+k_2+k) \,  K_{3}
\nonumber \\
& & \times\int {\rm d}\mu\ {\rm d}^2z_1\ {\rm d}^2z_2\ {\rm
d}^2z_3{\prod_{i<j}|a_{ij}|^2\over 64|y(z_1) y(z_2) y^2(z_3)|^2}
\nonumber \\
& & \times \exp\{-\sum_{i<j}k_i\cdot k_j\langle
X(z_i)X(z_j)\rangle\}
\nonumber\\
& = & -ig_c^2g_{_M}C_I(2\pi)^{10} \delta^{10}(k_1+k_2+k )\,
{K}_{3} \nonumber \\
& & \times\int {1\over T^6} {\prod _{i=1}^4{\rm d}^2a_i\over {\rm
d} V_{pr} |\prod_{i<j}a_{ij}|^2}\ {{\rm d}^2z_1\ {\rm d}^2z_2\
{\rm d}^2z_3\over 64\,
|y(z_1) y(z_2) y^2(z_3)|^2} \nonumber \\
& & \times\exp\{ \langle X(z_1)X(z_2)\rangle-\langle
X(z_1)X(z_3)\rangle - \langle X(z_2)X(z_3)\rangle\} ,
\label{oneloopa}
\end{eqnarray}
where we have used the relation $2 k_i\cdot k_j = (k_i+k_j)^2 -
k_i^2 -k_j^2$ to set $k_1\cdot k_2 =- k_1\cdot k=-k_2\cdot k= -1$
and the kinematic factor ${K}_{3}$ is given as follows (for the
left-part only):
\begin{equation}
{K}_{3L} = 6\alpha_{\mu\nu\rho}\epsilon_2^\mu\epsilon_1^\nu
k_2^\rho+\sigma_{\mu\nu}(k_1^\mu k_2^\nu
\epsilon_1\cdot\epsilon_2-k_1^\mu\epsilon_2^\nu\epsilon_1\cdot
k_2-\epsilon_1^\mu\epsilon_2^\nu -\epsilon_1^\mu k_2^\nu
k_1\cdot\epsilon_2) .
\end{equation}
By comparing the one-loop amplitude given in eq.~(\ref{oneloopa})
with the relevant expression in eq.~(\ref{4PointsFact}), we see
that there is a mismatch of one factor of $T_i$ and an extra
integration over insertion point of the vertex operators. This is
due to the fact that there is a  translation invariance for the
one-loop amplitude. So we can fix an insertion point and this
cancels a factor $T$. The precise relation between $G$ and
$\langle X(z_i) X(z_j) \rangle$ is:
\begin{equation}
\langle X(z_i) X(z_j) \rangle  = - G(z_i,z_j) -  \ln |
y(z_i)y(z_j)K^2| \, .
\end{equation}
(See also eqs.~(\ref{eq25})--(\ref{eq28}).) Taking this into
account, we have
\begin{eqnarray}
 & & \hskip -1cm \int {1\over T^6} {{\rm d}^2z_1\ {\rm d}^2z_2\
{\rm d}^2z\over
|y(z_1) y(z_2) y^2(z)|^2} \nonumber \\
& & \times\exp\{ \langle X(z_1)X(z_2)\rangle-\langle
X(z_1)X(z)\rangle - \langle X(z_2)X(z)\rangle\} \nonumber \\
& = &   \int {|K|^2 \over T^6} {{\rm d}^2z_1\ {\rm d}^2z_2\ {\rm
d}^2z\over
|y(z_1) y(z_2) y (z)|^2} \nonumber \\
& & \times\exp\{ -G(z_1,z_2) + G(z_1,z) + G(z_2,z) \} \nonumber \\
& = &   \int {|K|^8 \over T^6}  {\rm d}^2\tilde z_1\ {\rm
d}^2\tilde z_2\ {\rm
d}^2\tilde z \, \exp\{ -G(z_1,z_2) + G(z_1,z) + G(z_2,z) \} \nonumber \\
& = &  \int {|K|^6 \over T^5}  {\rm d}^2\tilde z_1  {\rm
d}^2\tilde z_2 \, \exp\{ -G(z_1,z_2) + G(z_1,z) + G(z_2,z) \} \nonumber \\
& = &  \int {|K|^2 \over T^5} {{\rm d}^2z_1\ {\rm d}^2z_2 \over
|y(z_1) y(z_2)  |^2} \nonumber \\
& & \times\exp\{ -G(z_1,z_2) + G(z_1,z) + G(z_2,z) \} .
\end{eqnarray}
This agrees with eq.~(\ref{4PointsFact}) for the relevant part of
the two one-loop amplitudes.

To complete the study of the factorization we would also like to
show that the kinematic factor $K(k_i,\epsilon_i)$ also factorizes
correctly. This was done in \cite{Yasuda,Zhu,Zhu0}. Here we will
give more details for the summation over intermediate tensors.
This factorization is true for both the left- and right-part.
Setting $\epsilon_i^{\mu\nu}=\epsilon_i^\mu \tilde\epsilon_i^\nu$,
we have
\begin{equation}
K(k_i,\epsilon_i) = K_L(k_i,\epsilon_i) K_R(k_i, \tilde\epsilon_i)
.
\end{equation}
In the following we discuss the factorization of $K_L$ only and by
a abuse of notation we simply write $K(k_i,\epsilon_i)$ for
$K_L(k_i,\epsilon_i) $.

Setting
\begin{eqnarray}
K_{3,1}(k_1,\epsilon_1,k_2,\epsilon_2,k,\alpha) & = &  6
{\epsilon_1}_\mu
{\epsilon_2}_\nu {k_1}_\rho \alpha^{\mu\nu\rho} \nonumber \\
K_{3,2}(k_1,\epsilon_1,k_2,\epsilon_2,k_3,\sigma) & = & -
\sigma^{\mu\nu}({{\epsilon_1}_{\mu }} {{\epsilon_2}_{\nu
}}+{{k_1}_{\mu }}
 {{\epsilon_2}_{\nu }} \epsilon_1\cdot k_2
 \nonumber \\
 & & +{{k_1}_{\mu }} {{k_1}_{\nu }}
 \epsilon_1\cdot \epsilon_2 -{{\epsilon_1}_{\nu }} {{k_1}_{\mu }}
 \epsilon_2\cdot k_1),
\end{eqnarray}
and by using the completeness relations eqs.~(\ref{completealpha})
and (\ref{completesigma}) for the polarization tensors, we have
 \begin{eqnarray}
& & \hskip -1cm
 \sum_\alpha K_{3,1}(k_1,\epsilon_1,k_2,\epsilon_2,k,\alpha)
 K_{3,1}(k_3,  \epsilon_3,k_4,\epsilon_4,-k,\alpha)\nonumber \\
 & = &-\frac{1}{2}\Big\{
 (u (\epsilon_1\cdot \epsilon_4 \epsilon_2\cdot \epsilon_3-
 \epsilon_1\cdot \epsilon_3 \epsilon_2\cdot \epsilon_4 -
 \frac{1}{2} \epsilon_2\cdot \epsilon_3 \epsilon_1\cdot k_2
 \epsilon_4\cdot k_3 \nonumber \\
 & & +\frac{1}{2}
 \epsilon_2\cdot \epsilon_4 \epsilon_1\cdot k_2
 \epsilon_3\cdot k_4
 - \frac{1}{2} \epsilon_1\cdot \epsilon_4 \epsilon_2\cdot k_1
\epsilon_3\cdot k_4  \nonumber \\
& & +\frac{1}{2}
 \epsilon_1 \cdot \epsilon_3
 \epsilon_2\cdot k_1 \epsilon_4\cdot k_3))
 +\epsilon_1\cdot \epsilon_4 \epsilon_2\cdot
\epsilon_3-\epsilon_1\cdot \epsilon_3 \epsilon_2\cdot \epsilon_4
 \nonumber \\
& & +\epsilon_2\cdot \epsilon_3 (-\epsilon_1\cdot k_4 \epsilon_4
 \cdot k_1-\epsilon_1\cdot k_3 \epsilon_4\cdot k_2 )  \nonumber \\
& & -\epsilon_2\cdot \epsilon_4 (-\epsilon_1\cdot k_4
 \epsilon_3\cdot k_1-
 \epsilon_1\cdot k_3 \epsilon_3\cdot k_2)
\nonumber \\
& & +\epsilon_1\cdot
 \epsilon_4 (-\epsilon_2\cdot k_4 \epsilon_3
 \cdot k_1-\epsilon_2\cdot k_3
 \epsilon_3\cdot k_2)  \nonumber \\
& &  -\epsilon_1\cdot \epsilon_3
 (- \epsilon_2\cdot k_4
 \epsilon_4\cdot k_1- \epsilon_2\cdot k_3
 \epsilon_4\cdot k_2)
\nonumber \\
& & +(\epsilon_1\cdot k_4 \epsilon_2\cdot k_3-\epsilon_1\cdot k_3
 \epsilon_2\cdot k_4)  \nonumber \\
& & \times  (\epsilon_3\cdot k_2
 \epsilon_4\cdot k_1-\epsilon_3
 \cdot k_1 \epsilon_4\cdot k_2) \Big\}\, ,
\end{eqnarray}
and
\begin{eqnarray}
& & \hskip -1cm \sum_\sigma
K_{3,2}(k_1,\epsilon_1,k_2,\epsilon_2,k,\sigma)
K_{3,2}(k_3,\epsilon_3,k_4, \epsilon_4,-k,\sigma) \nonumber
\\
 & = & \frac{1}{2}\left \{
 \frac{1}{2} {u^2}\epsilon_1 \cdot \epsilon_2 \epsilon_3 \cdot
 \epsilon_4 + u {[}\epsilon_3 \cdot \epsilon_4
(\epsilon_1 \cdot k_4 \epsilon_2 \cdot k_3 -\epsilon_1 \cdot k_3
 \epsilon_2 \cdot k_4 ) \nonumber \right.\\
& &
 + \epsilon_1 \cdot \epsilon_2
\epsilon_3 \cdot \epsilon_4+\epsilon_1 \cdot \epsilon_2(\epsilon_3
\cdot k_2 \epsilon_4
 \cdot k_1-\epsilon_3 \cdot k_1 \epsilon_4 \cdot
 k_2) \nonumber \\
& & +\frac{1}{2}\epsilon_1 \cdot \epsilon_4 \epsilon_2 \cdot k_1
 \epsilon_3 \cdot k_4   +\frac{1}{2}\epsilon_2 \cdot
 \epsilon_3 \epsilon_1 \cdot k_2 \epsilon_4 \cdot k_3
 \nonumber \\
 & & -\frac{1}{2}\epsilon_1 \cdot \epsilon_3
 \epsilon_2 \cdot k_1 \epsilon_4 \cdot k_3
 -\frac{1}{2}\epsilon_2 \cdot \epsilon_4 \epsilon_1 \cdot k_2
 \epsilon_3 \cdot k_4 {]}
 \nonumber \\
 & & +\epsilon_2 \cdot \epsilon_4
 (\epsilon_1 \cdot k_2 \epsilon_3 \cdot k_1
 +\epsilon_1 \cdot k_3 \epsilon_3 \cdot k_4)
 \nonumber \\
 & & +\epsilon_1 \cdot \epsilon_4
 (\epsilon_2 \cdot k_4 \epsilon_3 \cdot
 k_1 -\epsilon_2 \cdot k_3 \epsilon_3 \cdot k_2)
 \nonumber \\& &
 +\epsilon_2 \cdot \epsilon_3 (\epsilon_1
 \cdot k_3 \epsilon_4 \cdot k_2-\epsilon_1
 \cdot k_4 \epsilon_4 \cdot k_1)
\nonumber \\
 & &  +\epsilon_1 \cdot \epsilon_3 (\epsilon_2
 \cdot k_1 \epsilon_4 \cdot k_2+
 \epsilon_2 \cdot k_4 \epsilon_4 \cdot k_3)
 \nonumber \\
 & & -2 \epsilon_1 \cdot \epsilon_2
 \epsilon_3 \cdot k_1 \epsilon_4 \cdot
 k_2-2\epsilon_3 \cdot \epsilon_4 \epsilon_1 \cdot k_3
 \epsilon_2 \cdot k_4
\nonumber \\
 & &  +\epsilon_1 \cdot \epsilon_4 \epsilon_2 \cdot \epsilon_3
 +\epsilon_1 \cdot \epsilon_3 \epsilon_2 \cdot \epsilon_4
 \nonumber \\& &
 +(\epsilon_1 \cdot k_4 \epsilon_2 \cdot
 k_3-\epsilon_1 \cdot k_3 \epsilon_2 \cdot
 k_4) \nonumber \\
 & & \times (\epsilon_3 \cdot k_2
 \epsilon_4 \cdot k_1-\epsilon_3 \cdot k_1 \epsilon_4 \cdot
 k_2)\left. \right\} .
\end{eqnarray}
By using the above results one proves that the kinematic factor
$K(k_i,\epsilon_i)$ factorizes into a product of two one-loop
kinematic factors. Explicitly we have:
\begin{eqnarray}
 & & \hskip -1cm K(k_i,\epsilon_i)|_{(k_1+k_2)^2 = -2} = {u^2\over
4}\epsilon_1\cdot\epsilon_2\epsilon_3\cdot\epsilon_4+{u\over
2}\Big[\epsilon_1\cdot\epsilon_2\epsilon_3\cdot\epsilon_4+
\epsilon_1\cdot\epsilon_3\epsilon_2\cdot\epsilon_4
\nonumber \\
 & & -\epsilon_1\cdot\epsilon_4\epsilon_2\cdot\epsilon_3
 +  \epsilon_1 \cdot \epsilon_2(\epsilon_3 \cdot k_2 \epsilon_4
\cdot k_1-\epsilon_3 \cdot k_1 \epsilon_4 \cdot
 k_2) \nonumber \\
 & & +\epsilon_3 \cdot \epsilon_4
 (\epsilon_1 \cdot k_4 \epsilon_2
 \cdot k_3-\epsilon_1 \cdot k_3 \epsilon_2 \cdot k_4 )
 \nonumber \\
 & &+\epsilon_1 \cdot \epsilon_4
 \epsilon_2 \cdot k_1 \epsilon_3 \cdot k_4
 +\epsilon_2 \cdot \epsilon_3 \epsilon_1
 \cdot k_2 \epsilon_4 \cdot k_3
 \nonumber \\
 & & -\epsilon_1 \cdot \epsilon_3 \epsilon_2
 \cdot k_1 \epsilon_4 \cdot k_3
 -\epsilon_2 \cdot \epsilon_4 \epsilon_1
 \cdot k_2 \epsilon_3 \cdot  k_4\Big]
 \nonumber \\
 & & +\epsilon_1 \cdot \epsilon_3(\epsilon_2
 \cdot k_3 \epsilon_4 \cdot k_1-\epsilon_2
 \cdot k_1 \epsilon_4 \cdot  k_3)
 \nonumber \\
 & & +\epsilon_2 \cdot \epsilon_4(\epsilon_1
 \cdot k_4 \epsilon_3 \cdot
 k_2
 - \epsilon_1 \cdot k_2 \epsilon_3 \cdot k_4)\nonumber \\& &
 +\epsilon_2 \cdot \epsilon_3 \epsilon_1 \cdot k_3 \epsilon_4 \cdot
 k_2+\epsilon_1 \cdot \epsilon_4 \epsilon_2
 \cdot k_4 \epsilon_3 \cdot
 k_1-\epsilon_1 \cdot \epsilon_2 \epsilon_3
 \cdot k_1 \epsilon_4 \cdot
 k_2 \nonumber \\
 & & -\epsilon_3 \cdot \epsilon_4 \epsilon_2
 \cdot k_4 \epsilon_1 \cdot
 k_3 +\epsilon_1 \cdot \epsilon_3\epsilon_2 \cdot \epsilon_4
 \nonumber \\
 & = &\sum_\alpha K_{3,1}(k_1,\epsilon_1,k_2,\epsilon_2,k,\alpha)
 K_{3,1}(k_3,
 \epsilon_3,k_4,\epsilon_4,-k,\alpha) \nonumber \\
& &   +  \sum_\sigma
K_{3,2}(k_1,\epsilon_1,k_2,\epsilon_2,k,\sigma)
K_{3,2}(k_3,\epsilon_3,k_4, \epsilon_4,-k,\sigma) .
\end{eqnarray}
with ${k=-(k_1+k_2)}$. This completes the study of the
factorization of the two-loop four-particle amplitude in
superstring theory.

One could use the above result to fix the overall factor of the
two-loop four-particle amplitude. In order to do this we need the
precise overall factor for the one-loop three-particle amplitude.
This factor is unknown in the literature and it should be
determined by either using unitarity or factorization at one-loop.
We will leave these for a future publication \cite{XiaoZhu}.

Note added: Recently the 2-loop 4-particle amplitude in
superstring theory was also obtained in \cite{Phong3, Phong4} and
its factorization was studied by D'Hoker, Gutperle and Phong in
\cite{Phong5}.

\end{document}